# "Mechanically prepared copper surface in oxidizing and non-oxidizing conditions"

Daria Serafin*[1], Wojciech J. Nowak[1], Bartek Wierzba[1]

[1] Department of Materials Science, Faculty of Mechanical Engineering and Aeronautics, Rzeszow University of Technology, al. Powstancóv Warszawy 12, 35-959 Rzeszów, Poland

*corresponding author: Daria Serafin, e-mail address: darseraf@gmail.com, phone: +48 665 772 816

**Abstract.**

In the present work surface roughness of copper samples after mechanical surface preparation processes (polishing, grinding, sand-blasting) and after exposure in non-oxidizing (high purity argon) and oxidizing (air) conditions at 700 and 750°C have been evaluated by contact profilometer and by fractal analysis. Quantitative description of surfaces that are hidden from the sight of conventional methods is possible by fractal analysis, e.g. in the present work, the roughness of $Cu_2O$/Cu boundary has been evaluated under oxide layer. It has been pointed out, that in both oxidizing and non-oxidizing conditions initial surface roughness of copper is affected by annealing at high temperature: at 700°C surface roughness of polished and ground samples increased but at 750°C decreased in comparison to surface roughness prior to heat treatment. For sand-blasted samples in both oxidizing and non-oxidizing conditions the lowest values of surface roughness parameters were obtained at 750°C. It is then concluded that two phenomena are responsible for such effect: chemical reaction between copper and oxygen and surface diffusion of copper due to high temperature of annealing.

# 1. Introduction

Copper is widely used as an interconnector in ultralarge-scale integration devices, due mainly to its excellent electric conductivity [1–3]. Moreover, copper oxide is characterized by an appropriate band gap for using it, e.g. as solar cells [4–6]. However, in most of applications copper must have stable, oxide-free surface, which is not easy to achieve in practice [3]. Development of the methods, that may enable for improving the stability of surfaces require better understanding of the behavior of copper at high temperatures [2].

Processes, that occur on the copper surface change, when material is exposed to oxygen-rich environment. Proceeding reactions at high temperature appear in the following order: oxygen chemisorptions on the clean metal surface, nucleation of the oxide and its growth. Everything is followed by bulk oxide growth [5]. Oxidation of copper is an effect of outward diffusion mechanism of Cu cations via vacancies. The product of oxidation process is $Cu_2O$ oxide, which may further oxidize to CuO oxide [7–9]. It is well established, that above 600°C oxidation process proceeds mostly due to lattice diffusion. At lower temperatures, a great contribution to the overall oxidation rate makes grain boundary diffusion, which enables copper or oxygen atoms to diffuse easily through the oxide layer [9]. Different ratio between oxide thicknesses ($Cu_2O$ to CuO) obtained after oxidation of the sample of the same material in the same conditions means, that different diffusion mechanism is observed, due to different number of defects and grain boundary in $Cu_2O$ layer [10,11]. However, it is not known, how different number of defects and grain boundaries on the surface of pure copper affects the number of defects in $Cu_2O$ layer and therefore the kinetics of oxide growth. So, even though oxidation of copper in air has been extensively studied by many authors [12–14], there are difficulties in comparison of the results obtained in particular studies due to different purity of copper, its surface roughness and number of defects present in the near-surface region. Also the temperature, humidity, oxygen partial pressure, grains orientation and presence of impurities may significantly affect the oxide growth, especially in the initial stages of oxidation [5].

Mechanical surface preparation processes have a great influence on surface roughness and the state of the material. For instance, it has been proved, that diffusion rate of chromium in stainless steel may be enhanced by ball milling as a result of grain refinement and increased number of grain boundaries [15]. Moreover, surface roughness tends to have an influence on oxidation rate of nickel superalloys [16]. Oxidation resistance of pure copper is also affected by increased surface roughness and number of defects. Therefore, surface preparation process of components made out of copper is of prime importance in further applications [17,18].

Influence of polishing, grinding and sand-blasting processes on oxidation kinetics of copper samples at 650-750°C has been studied in our previous work [18]. It had been noticed that the boundary between sand-blasted copper and its oxide smoothed significantly after exposure for 2 h at 750°C. The authors proposed, that the effect may be connected with simultaneous inward and outward diffusion of oxygen and copper, respectively. However, the smoothening of copper surface may occur as well as the consequence of surface diffusion of copper at high temperature [19] or recrystallization processes [20].

Thus, the aim of this paper is to study changes in surface roughness of polished, ground and sand-blasted copper samples between cooper and its oxide after exposure for 2 h in air at 700 and 750°C. Then, comparison of the results with roughness of the same prepared samples after exposure for 2 h in an inert gas (argon) was performed. Therefore, a relationship between surface roughness and copper's behavior at high temperature was established. Roughness had been described mostly by fractal analysis method which is an easy way of evaluation of surface roughness under oxide layer. Such analysis may not be performed by contact profilometer without destroying the oxides.

## 2. Material and methods

### 2.1. Surface preparation

Block of high purity copper (99,9999 at. % Cu) had been cut into rectangular samples with the dimensions of approximately 13 x 14 x 2 mm. The small hole (2 mm in diameter) was drilled close to the center of top edge of all of the samples to hang them in the furnace (Fig. 1). Samples were then prepared in three different processes: polishing, grinding and sand-blasting. First sample had been ground mechanically using SiC papers with increasing gradation of the grain from 220 grit to 1200 grit. Polishing was performed using woven clothes with decreasing gradation of $SiO_2$ suspension, starting from 6 μm and finishing on 1 μm. Ground samples were prepared on SiC paper with the grain size of 220, while sand-blasted samples were initially ground on that paper and then sand-blasted with alumina ($Al_2O_3$) particles within compressed air (0,8 MPa). After mechanical preparation of the surface and prior to heat treatment under different conditions, samples were cleaned in ethanol using ultrasonic cleaner. One set of samples (polished, ground, sand-blasted) was used for surface roughness evaluation. The macrophotograps of the samples after these steps are presented in Fig. 1.

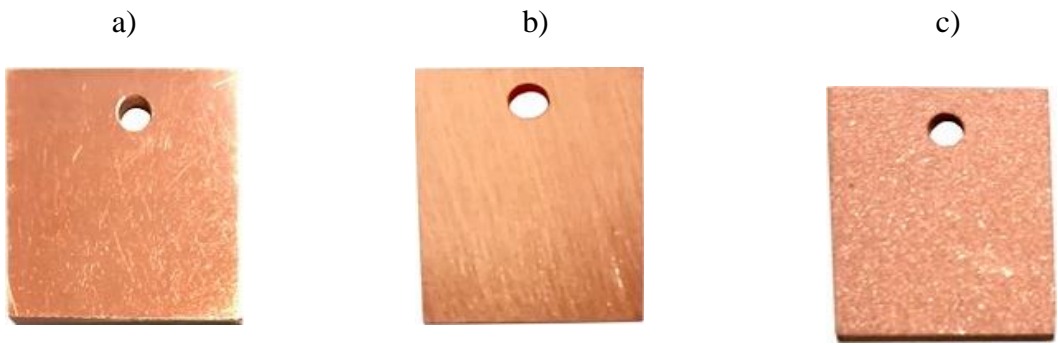

Fig. 1. Macrophotographs of samples after surface preparation treatments: a) polishing, b) grinding, c) sand-blasting.

**2.2. Surface roughness evaluation**

Surface roughness of the substrate has a great influence on the adhesion of thermally sprayed coatings and therefore on their lifetime [21–23], corrosion resistance of many materials [16,24], in diffusion bonding [25] and tribological behavior of the surfaces [22,26]. Accuracy in evaluation of surface roughness and its topography is of great importance in those applications. Therefore, there are many methods that enable for characterization and analysis of surface roughness. Conventional, and most often used one, that provides quantitative data about surface roughness bases on using traditional contact profilometer. In this method, the tip of measuring needle must remain in contact with the characterized surface during whole process. It may cause difficulties in reaching all the irregularities of convoluted surfaces as some of the deep valleys may be hidden under some overhangs. This is one of the most crucial limitations of contact profilometers [27,28].

Fractal analysis is one of the promising methods of describing surface roughness in quantitative way. Many different software are used for surface roughness evaluation by this method. All of them base on digital image analysis. It ensures that no deep valley on convoluted surface will be left "blind" during analysis [29].

The most often used fractal, quantitative parameter is fractal dimension, which may describe morphology of the surface, its roughness and texture [29,30]. Even though, fractal evaluation of surface roughness has its limitations, mostly connected with repeatability of image acquisition and processing of image adjustments, it is more often used as a method of evaluation of surface roughness as it gives more information about surface complexity [30].

### 2.2.1. Contact profilometer

The most popular amplitude parameter that enables for comparison of surface roughness of different samples is arithmetic average height $R_a$. It may be calculated by the following equation:

$$R_a = \frac{1}{n}\sum_{i=1}^{n}|y_i| \tag{1}$$

where n – number of intersection of the profile at the mean line, $y_i$ – the vertical distance from the mean line [31].

In the present work, surface roughness had been measured conventionally by Hommel Werke T8000 profilometer. For each sample 5 different measurements had been performed, which enabled for calculating average value and standard deviation of the results. The direction of measurement was indifferent for polished and sand-blasted samples, as their surface topography was isotropic. In case of ground sample, the measurement was carried out perpendicularly to the grinding scratches. The conditions of measurements were as follows: traverse set – 4 mm, linear speed – 0.5 mm/s.

### 2.2.2. Fractal analysis

Fractal theory has been recently used to investigate irregular surfaces, e.g. for evaluation of fractal roughness in nanoscale grinding [32], turning [33] and quantification of metallic corrosion [30,34]. This mathematical method bases on possibility of describing self-similar objects, by using one of a scales, that may be defined by apparent length of a surface profile. Such scale depends only on the size of the apparatus, which is used for measuring the scale. Relative length $L_R$ can be calculated by following equation:

$$L_R = \frac{\sum_{i}^{N} \frac{1}{\cos\theta_i} p_i}{L} \tag{2}$$

where $L$ – total projected length of all N virtual steps, $p_i$ – projected length of 1 virtual step, $\theta_i$ - angle between two surfaces: normal and nominal.

Examples of evaluation of $L_R$ parameter have been shown in Fig. 2a ($r$ – scale length). The length of virtual ruler must be constant in each calculation. Virtual steps between sampling intervals are located by linear interpolation. Relation between relative length ($L_R$) and scale length ($r$) enables for preparation of plots, that are used for comparison of roughness of different materials (Fig. 2b) [33,35].

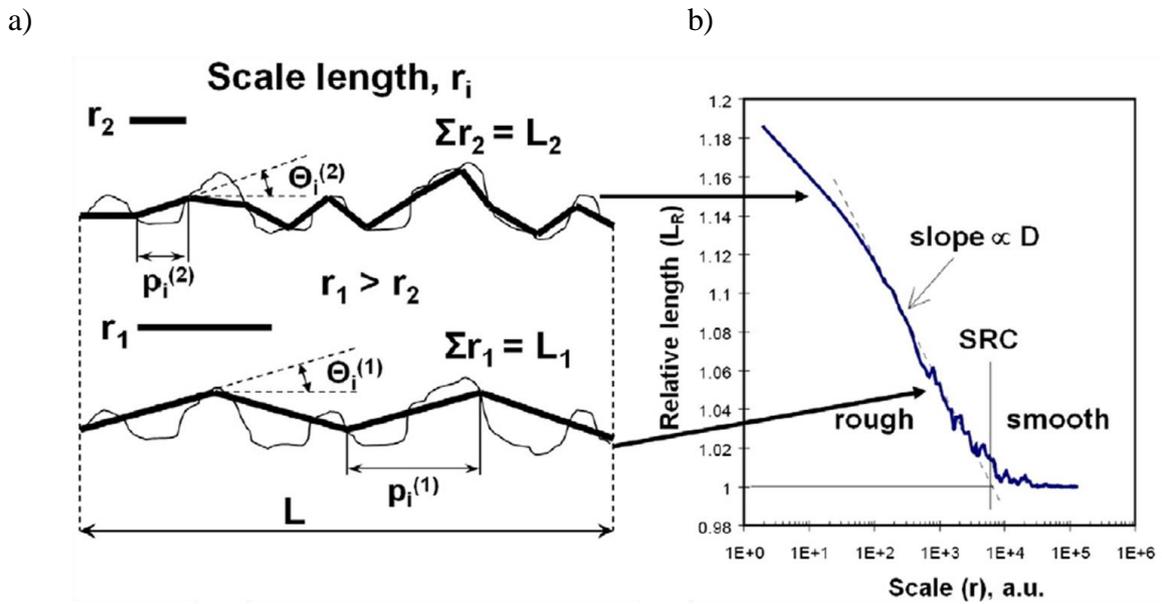

Fig. 2. Schematic example of a) method of evaluation of surface roughness using different ruler length, b) plot of relative length in the function of scale [35].

The fractal parameters, that describe the surface roughness, e.g. fractal dimension $FD$ can be evaluated from plots of relative length in the function of scale (Fig. 2b). Thus, fractal dimension is defined as:

$$FD = 1 + |slope| \qquad (3)$$

Correlation between fractal dimension parameter and the slope of relative length $L_R$ vs. scale $r$ plot means, that the more convoluted the surface is, the highest value of $FD$ is obtained [35].

Roughness analysis by fractal method requires performing a few steps [29,35]:

- acquisition of microphotograph of metallographic cross-section using optical microscope,
- conversion of microphotographs into binary scale images (metallic substrate – white, mounting and oxide layer – black),
- digitalization of converted microphotographs into two-dimentional x-y data sets, characterized by constant x step,
- run of analysis of data sets by commercially available software (in this work Sfrax 1.0. software [36] was used),
- evaluation of obtained results.

In this paper, 5 microphotographs were taken at different places on the sample, which were then analyzed according to the above mentioned procedure. The average value of surface roughness parameters were calculated for each sample.

## 2.3. Heat treatment

After surface preparation two sets of the samples had been subjected to high temperature exposure in non-oxidizing and oxidizing conditions for 2 h (one set of samples was composed out of three samples: polished, ground and sand-blasted).

### 2.3.1. Oxidizing process

Oxidation process had been carried out in thermogravimetrical furnace Xerion in air at 700°C and 750°C. Samples were hanged inside the glass tube in the furnace (one sample at a time) on the kanthal wire. Furnace had to be preheated to the required temperature, prior to oxidation test. After reaching the required value, the heating part of the furnace was moved from cooling to heating position. The samples had been taken out of the furnace and cooled to the room temperature after 2 h of oxidation.

**2.3.2. Non-oxidizing conditions**

The annealing process was carried out at 700 and 750°C in argon atmosphere - prior heating the samples were flushed with high purity argon (5.0) for 2 h to evacuate remainders of air out of the furnace. Samples had been then transported into hot part of the furnace, preheated earlier to 700 and 750°C, respectively. During whole process the inert gas (high purity argon) was used for protection of copper against oxidation. After 2 h of heating, samples were moved back to the cold section of the furnace and left there for cooling to the room temperature.

**2.4. Analysis and observations**

Prior to heat treatment processes, the surface roughness of the samples had been evaluated using two methods – conventional contact profilometer and fractal analysis method. After the oxidation process the samples were sputtered with thin gold coating for oxide protection and for increasing the contrast between the formed oxide scale and resin during microscopic observations. Surface roughness of the samples after exposure in argon were additionally measured by conventional contact profilometer.

After heat treatment, metallographic cross-sections of the samples had been prepared and observed using scanning electron microscope Hitachi S3400 (SEM) and optical microscope Nikon Epiphot 300. The microphotographs obtained using optical microscope were also the base for fractal analysis of the surface roughness after heat treatment.

**3. Results**

**3.1. Surface roughness prior to heat treatment**

Mechanically prepared surface of copper samples had been examined in terms of its roughness. Firstly, conventional contact profilometer had been used. Arithmetic average height parameter ($R_a$) has shown the great differences between the samples: polished one has been characterized by the lowest surface roughness, almost 70 times lower than sand-blasted one and 30 times lower than ground sample. Comparison of results obtained by conventional contact profilometer are presented in Fig. 3 and in Table 1.

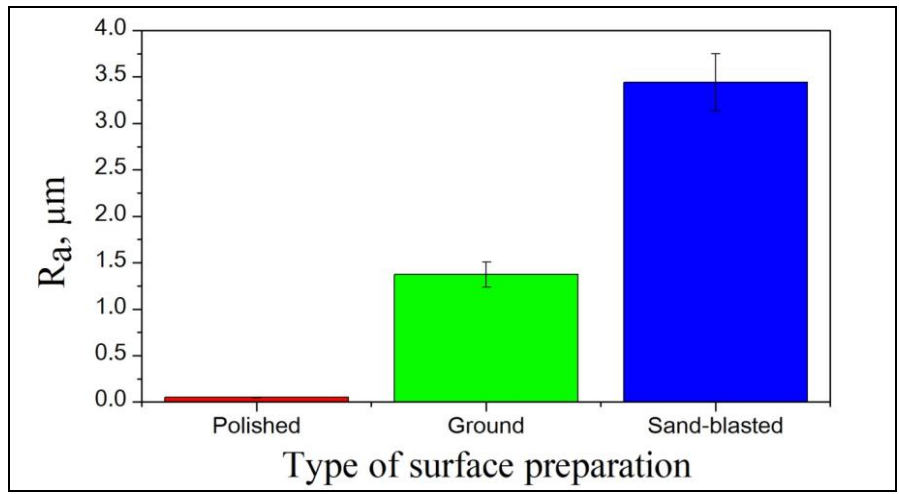

Fig. 3. Surface roughness of samples after different type of mechanical preparation obtained by contact profilometer.

Table 1. Differences in $R_a$ parameter for samples after surface preparation processes measured by contact profilometer.

| Cu | Polished | Ground | Sand-blasted |
|---|---|---|---|
| $R_a$, μm | 0.050 | 1.373 | 3.440 |
| Standard deviation SD | 0.003 | 0.136 | 0.306 |

Surface roughness had been also analyzed using fractal analysis. Procedure of evaluation of surface roughness by this method resulted firstly in preparation of binary scale images of surface profiles (Fig. 4), then the procedure described in chapter 2.2.2 was followed and consequently plots of relative length $L_R$ as a function of relative scale *r* were produced (Fig. 5). The results confirmed the differences in roughness of the surfaces obtained during different surface preparation processes. The fractal parameters are presented in Table 2.

a)                      b)                      c)

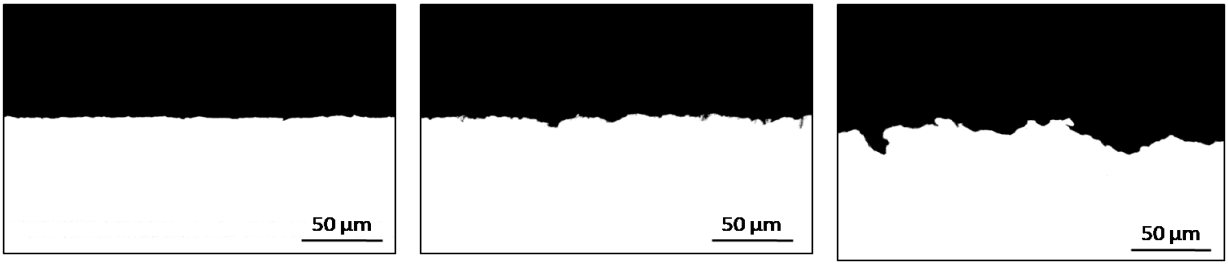

Fig. 4. Binary representation of surface profile in as-prepared condition: a) polished, b) ground, c) sand-blasted sample.

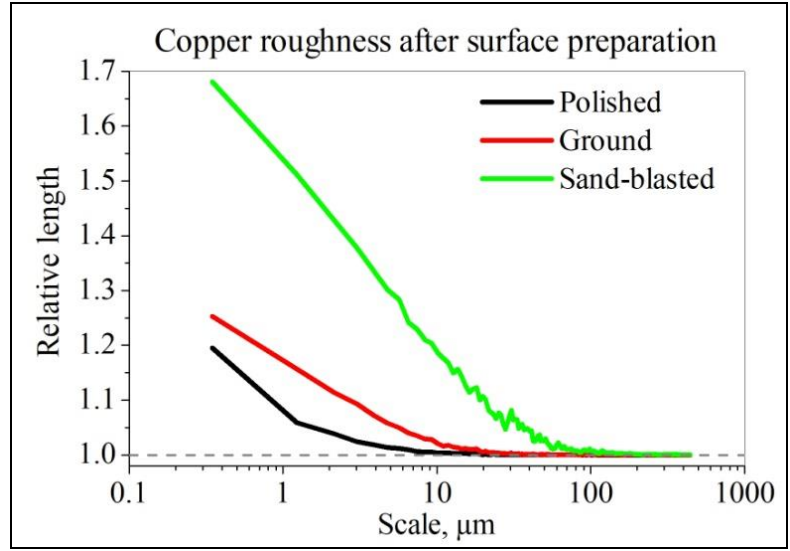

Fig. 5. Plot of relative length vs. scale for copper after surface preparation.

Table 2. Values of fractal parameters obtained for samples after surface preparation processes calculated by Sfrax 1.0 software.

| Parameter | Polished | Ground | Sand-blasted |
|---|---|---|---|
| **Fractal Dimension FD** | 1.013 | 1.037 | 1.110 |
| **SD** | 0.001 | 0.008 | 0.035 |
| **Relative length at 5 μm scale length ($L_R$ =5 mm)** | 1.012 | 1.054 | 1.293 |
| **SD** | 0.003 | 0.015 | 0.128 |

**3.2. Results of surface analysis after oxidation process**

The polished sample after exposure in air for 2 h at 700°C presented lower mass gain, than polished sample oxidized at 750°C (Fig. 6). The differences in mass gain was not significantly visible for ground samples, while sand-blasted samples demonstrated opposite trend to polished sample – the sand-blasted one oxidized at 750°C presented 2.5 times greater mass gain that the one oxidized at 700°C [18].

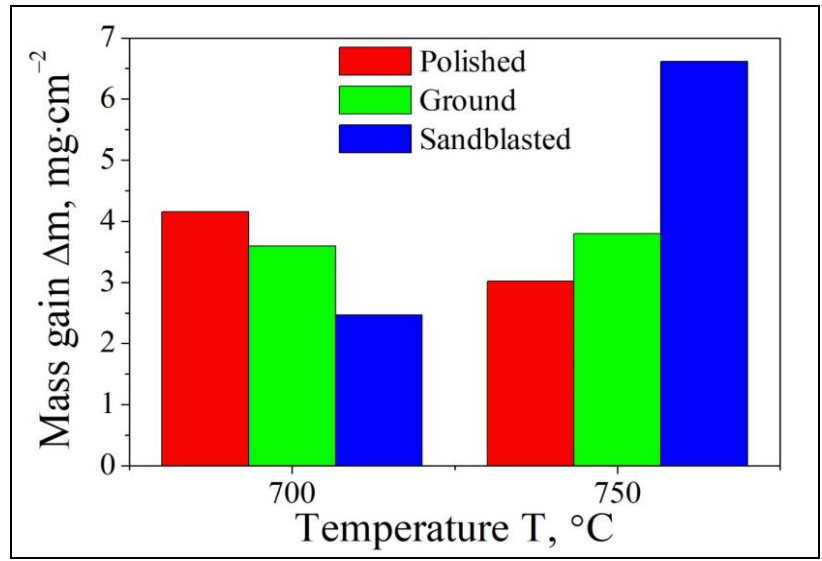

Fig. 6. Mass gain of polished, ground and sandblasted samples after oxidation at 700°C and 750°C in air for 2 h (reproduced from Ref. [18]).

The significant differences in oxidation mechanism of sand-blasted samples at 700 and 750°C can be noticed. At lower temperature, the mixed Cu/Al oxide was formed just near the oxide scale/copper interface, due to contamination of surface with alumina particles coming from sand-blasting process (Fig. 7c). The layer of $Cu_2O$ oxide of significant thickness had grown on the top of Cu/Al oxide, which is further covered with thin layer of CuO oxide. However, no mixed Cu/Al oxide was observed at higher temperature, but the layer of $Cu_2O$ oxide is much thicker than that observed on other samples (Fig. 7f). Moreover, it is visible on the microphotographs, that $Cu_2O$ grains has a columnar morphology on all of the polished and ground samples. On the contrary, on the sand-blasted sample oxidized at 750°C, round and small grains of $Cu_2O$ close to copper/$Cu_2O$ boundary are observed, which are then covered with columnar $Cu_2O$ oxide grains (Fig. 7f). Moreover, roughness of this sample also decreased after oxidation [18]. These observations forced the authors for further examination of surface roughness of the samples after oxidation.

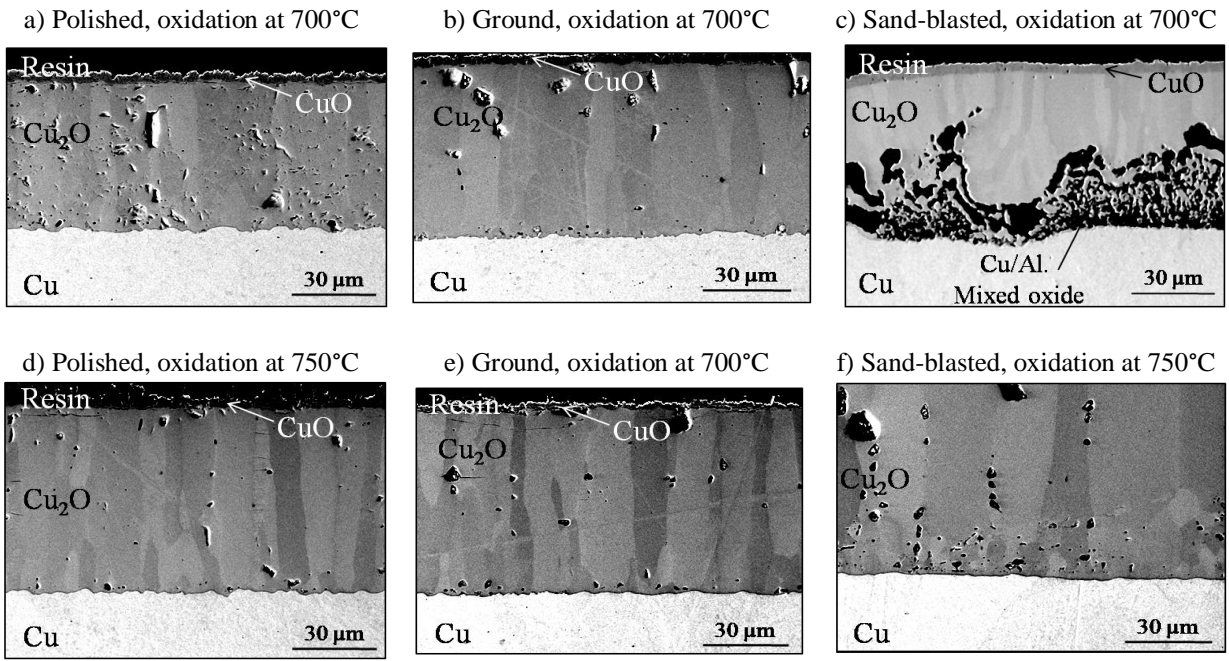

Fig. 7. SEM images of metallographic cross-sections of copper samples after oxidation in air for 2 h at different temperatures.

Binary scale images for fractal analysis were prepared in such a way, that white color substituted copper substrate and black – mounting and oxides (Fig. 8). Comparison of these photographs with binary scale images of the samples after surface preparation processes (Fig. 4) has shown vividly a great difference in surface profiles – the higher temperature of oxidation, the greater smoothening of the surface of sand-blasted surface was observed (Fig. 4c, Fig. 8c, f).

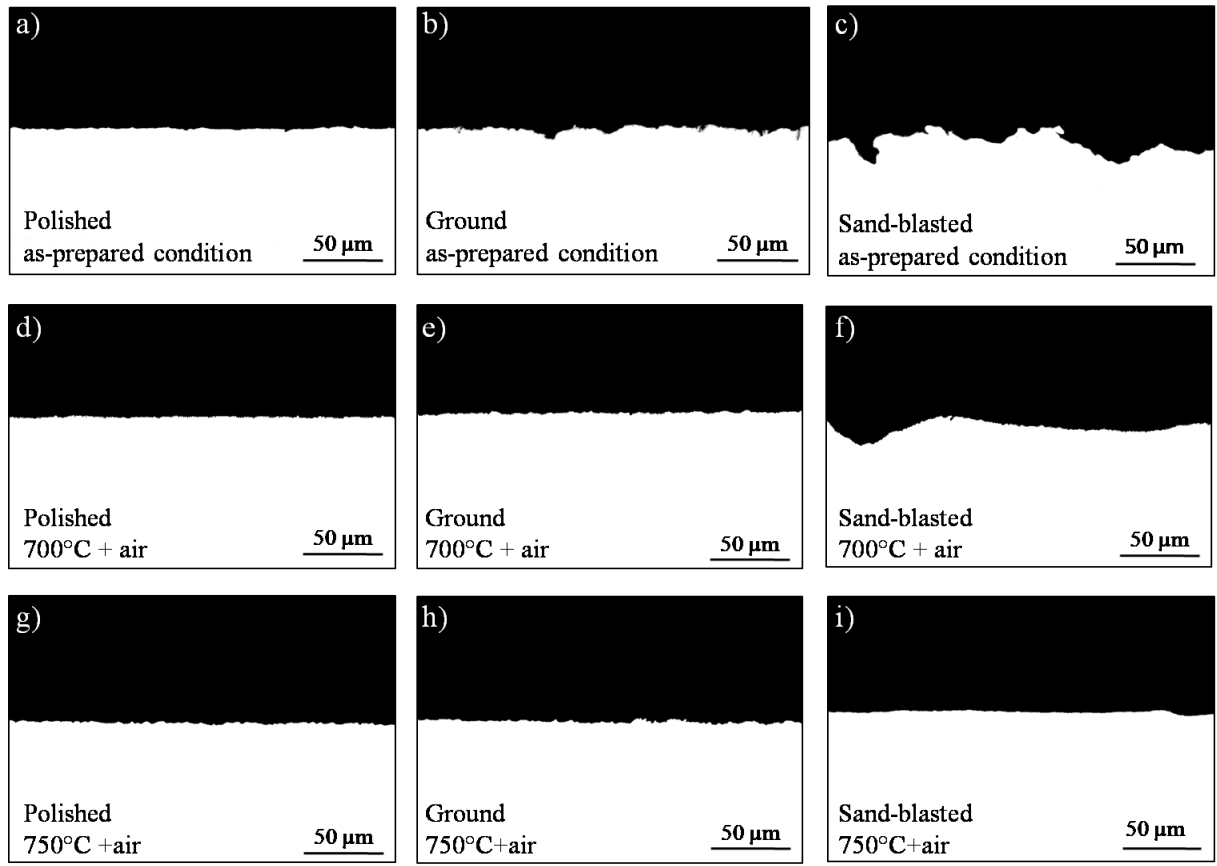

Fig. 8. Binary representation of surface profile in as-prepared condition (a-c), after oxidation at 700°C (d-f) and 750°C (g-i) for 2 h (white color – copper substrate, black – mounting and oxide layer).

Values of fractal dimension (FD) and Relative length ($L_R$) at 5 μm are presented in Table 3. The comparison of the relative length vs. scale plots after surface preparation and below the oxide layer after oxidation process is shown in Fig. 9.The differences in fractal parameters of surfaces under oxide layer are not visible for polished samples. However, at lower temperature (700°C) the average surface roughness of copper was higher below the oxide layer than before oxidation process (Fig. 9a). The decrease of roughness after exposure at 750°C was observed when compared to the initial value (in the as-prepared conditions). Similar observations were made for ground sample, but the plot of relative length vs. scale obtained for sample after oxidation at 700°C prevailed only at low values of scale length (Fig.

9b). However, there is visible decrease in fractal dimension and relative length values after oxidation for ground sample. Nevertheless, there is no significant difference between values of these parameters depending on the temperature of oxidation.

Temperature of oxidation has influence on roughness of copper surface under oxide layer for sand-blasted samples. For these samples, meaningful decrease in values of FD and $L_R$ parameters (Table 3) as well as in the slope of $L_R$ vs $s$ plot (Fig. 9c) between samples oxidized at different temperatures was observed. The lowest surface roughness was shown for sand-blasted sample oxidized at the highest temperature (750°C).

Table 3. Differences in fractal parameters for samples after surface preparation (SP) and after oxidation for 2 h.

|  | **Polished** | | | **Ground** | | | **Sand-blasted** | | |
|---|---|---|---|---|---|---|---|---|---|
|  | **After SP** | **Air at 700°C** | **Air at 750°C** | **After SP** | **Air at 700°C** | **Air at 750°C** | **After SP** | **Air at 700°C** | **Air at 750°C** |
| **FD** | 1.013 | 1.013 | 1.013 | 1.037 | 1.013 | 1.012 | 1.110 | 1.013 | 1.007 |
| **SD** | 0.001 | 0.003 | 0.003 | 0.008 | 0.002 | 0.001 | 0.035 | 0.003 | 0.001 |
| **$L_R$ =5 mm** | 1.012 | 1.011 | 1.010 | 1.054 | 1.010 | 1.009 | 1.293 | 1.029 | 1.006 |
| **SD** | 0.003 | 0.004 | 0.001 | 0.015 | 0.002 | 0.003 | 0.128 | 0.012 | 0.003 |

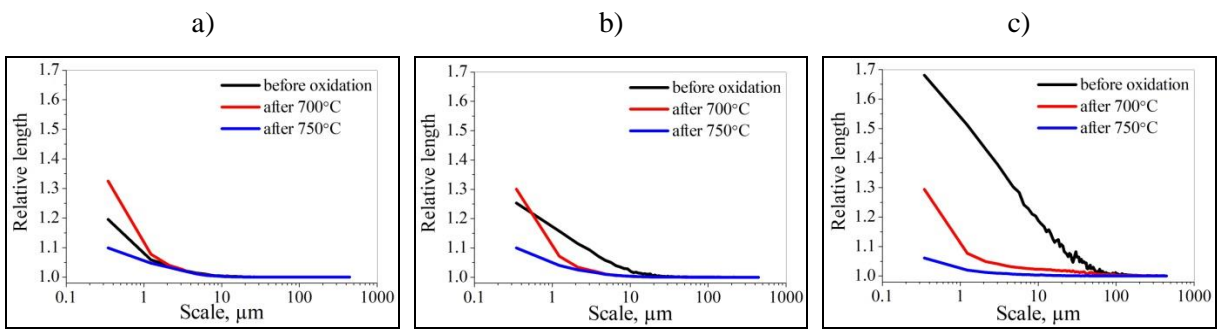

Fig. 9. Plots of relative length vs. scale for copper after surface preparation and under oxide layers for a) polished, b) ground and c) sand-blasted samples.

### 3.3. Results of exposure in an inert gas

The smoothening of Cu/Cu$_2$O boundary could result from surface diffusion of copper due to exposure of the samples at high temperature. Thus, additional exposure of the samples in an inert gas were performed at the same temperatures and the same time as oxidation test. Macrophotographs of the samples after annealing in an inert atmosphere are presented in Fig. 10.

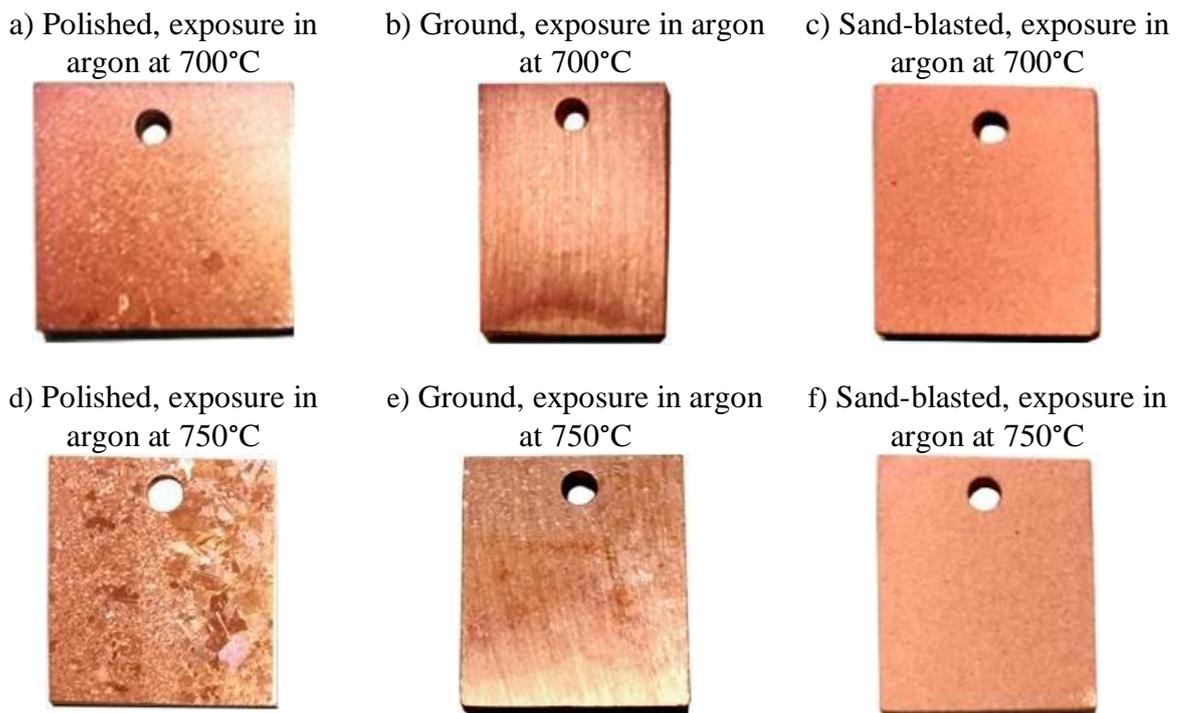

Fig. 10. Macrophotographs of the samples after exposure for 2 h in argon at 700 and 750°C of polished (a and d), ground (b and e) and sand-blasted (c and f) surfaces of copper.

The greatest difference was observed for polished sample after exposure at 700°C in argon. The small, round and regular recrystallized grains (Fig. 10a) are visible. Annealing of polished sample at 750°C resulted in further recrystallization process (Fig. 10d). The surface is covered with recrystallized grains, which are bigger and more irregular than these observed at lower temperature. Less differences are visible on ground samples (Fig. 10b, e). The macrostructure of these samples did not change significantly and the scratches after surface preparation in grinding process are still visible. Surface of sand-blasted samples were also not remarkably affected by high-temperature exposure in argon (Fig. 10c, f).

The surface roughness results obtained by the contact profilometer, compared to the results after surface preparation processes (Fig. 3, Table 1) are presented in Table 4 and in Fig. 11 (y-axis scale is different for polished, ground and sand-blasted samples).

Table 4. Comparison of differences in $R_a$ parameter for samples after surface preparation processes (SP) and after exposure in argon measured by contact profilometer.

|  | Polished | | | Ground | | | Sand-blasted | | |
| --- | --- | --- | --- | --- | --- | --- | --- | --- | --- |
|  | After SP | Argon at 700°C | Argon at 750°C | After SP | Argon at 700°C | Argon at 750°C | After SP | Argon at 700°C | Argon at 750°C |
| Ra, um | 0.050 | 0.106 | 0.053 | 1.373 | 1.338 | 1.326 | 3.440 | 3.166 | 3.105 |
| SD | 0.003 | 0.040 | 0.013 | 0.136 | 0.085 | 0.099 | 0.306 | 0.108 | 0.106 |

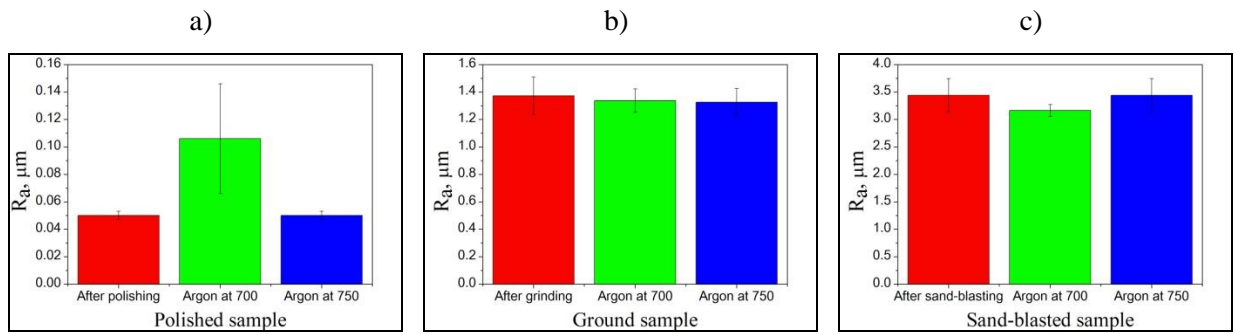

Fig. 11. Comparison of surface roughness of a) polished, b) ground and c) sand-blasted samples before and after exposure in argon at 700°C and 750°C for 2 h.

The differences in $R_a$ parameter are not statistically substantial for ground and sand-blasted samples. However, polished sample that was exposed in argon at 700°C revealed almost a double increase in $R_a$ parameter in comparison to others. This effect may be connected with proceeding recrystallization process and surface diffusion of cooper.

Fractal analysis of the samples exposed in argon based on the binary scale images presented in Fig. 12. The comparison of these scales images with the images in as-prepared condition shows, that the profiles of polished and ground surfaces are slightly more convoluted after heat treatment.

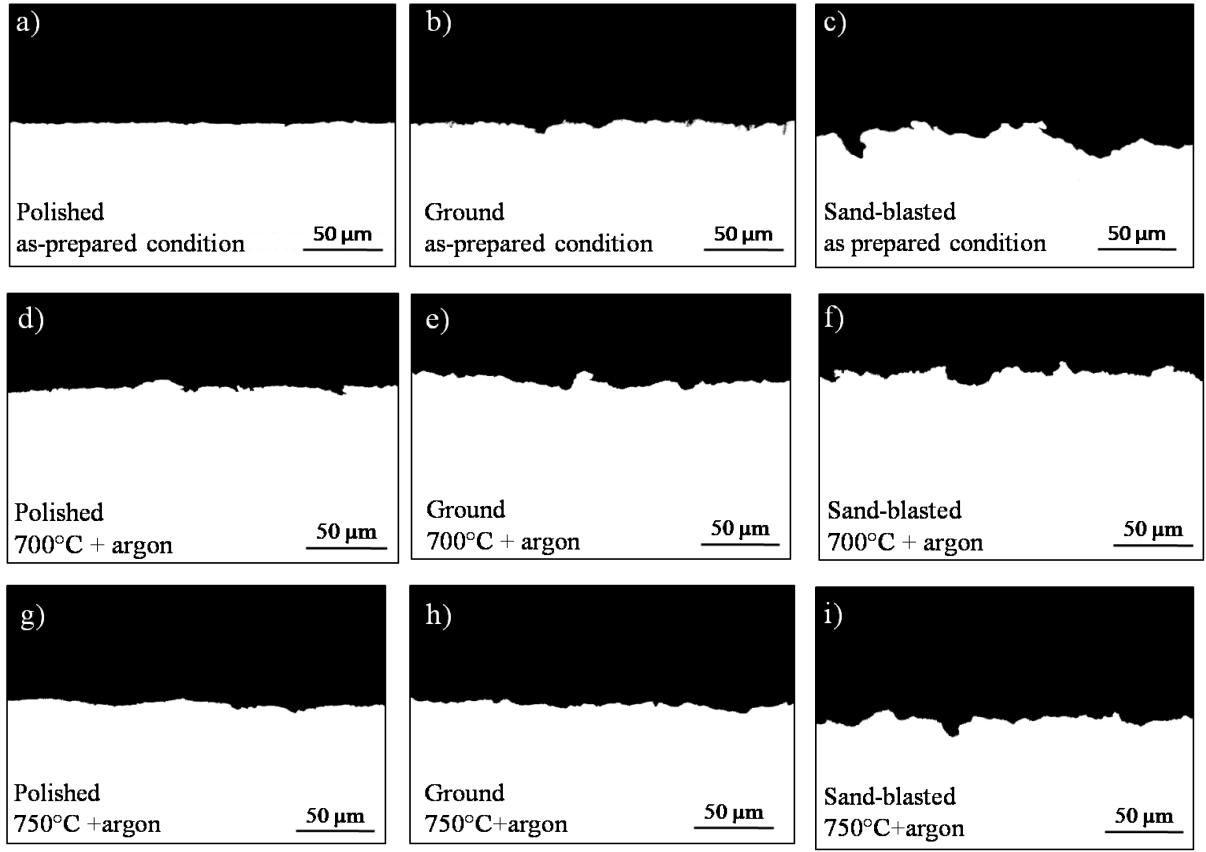

Fig. 12. Binary representation of surface profile in as-prepared condition (a-b) and after exposure in argon at 700°C (d-f) and 750°C (g-i) for 2 h (white color – copper substrate, black – mounting and oxide layer).

Moreover, the fractal parameters were established (Table 5) and plots of relative length vs. scale were prepared (Fig. 13). The differences in slope of the plots of $L_R$ vs. *r* are not significant for polished and ground samples. However, the slope for the samples exposed at the 750°C is in both situations lower than for others.(Fig. 13a, b). This trend may be also visible for sand-blasted sample, where the lowest value of $L_R$ was observed at the same scale *r* for sample after exposure in argon at 750°C and significantly the highest for the sample after surface preparation processes (Fig. 13). This information finds its confirmation in fractal

parameters – both FD and $L_R$ parameters has the lowest value for sample exposed in argon at 750°C and the highest – after surface preparation (Table 5). Similar observations may be made for FD parameter of ground sample, even though the differences between particular samples are not high. However, for polished sample, the highest value of fractal parameters is observed for sample exposed in argon at 700°C (Table 5). This observation is in good agreement with the value of parameter $R_a$, calculated by contact profilometer (Table 4).

Table 5. Differences in fractal parameters for samples after surface preparation (SP) and after exposure in pure argon for 2 h.

|  | Polished | | | Ground | | | Sand-blasted | | |
|---|---|---|---|---|---|---|---|---|---|
|  | After SP | Argon at 700°C | Argon at 750°C | After SP | Argon at 700°C | Argon at 750°C | After SP | Argon at 700°C | Argon at 750°C |
| FD | 1.013 | 1.020 | 1.011 | 1.037 | 1.034 | 1.025 | 1.110 | 1.059 | 1.052 |
| SD | 0.001 | 0.008 | 0.005 | 0.008 | 0.005 | 0.007 | 0.035 | 0.009 | 0.012 |
| $L_R$ = 5 mm | 1.012 | 1.031 | 1.018 | 1.054 | 1.065 | 1.045 | 1.293 | 1.129 | 1.109 |
| SD | 0.003 | 0.016 | 0.011 | 0.015 | 0.011 | 0.016 | 0.128 | 0.017 | 0.028 |

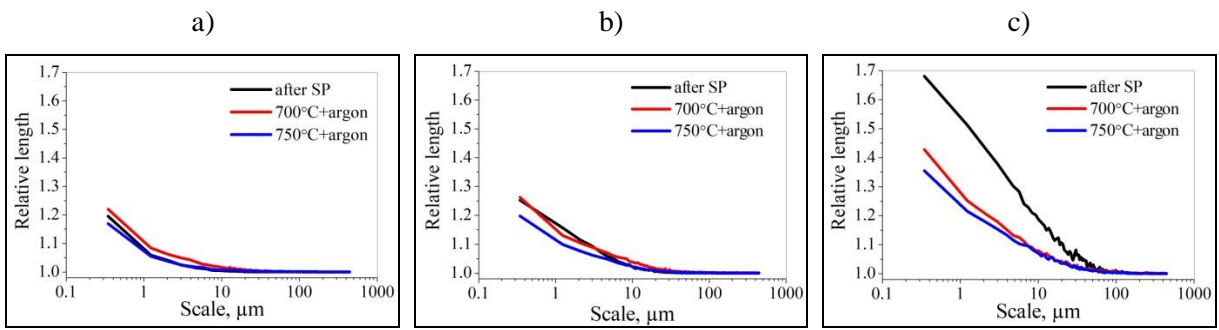

Fig. 13. Plots of relative length vs. scale for copper after surface preparation (SP) and after exposure for 2 h in pure argon for a) polished, b) ground, c) sand-blasted sample.

## 4. Discussion

Depending on initial roughness of copper and temperature of oxidation, the exposure of copper samples in oxidizing environment has a different influence on the Cu/Cu$_2$O boundary.

The roughness after oxidation was higher than in as-prepared condition at lower temperature for polished and ground samples (Fig. 9a, b). The significant drop in slope of relative length vs. scale plot was observed after oxidation at 700°C (Fig. 9c) for sand-blasted sample. The lowest roughness was observed after oxidation at 750°C for all of the samples.

Decreasing roughness of $Cu/Cu_2O$ interface may be caused by two main factors: chemical reaction between oxygen and copper or surface diffusion of copper atoms. The surface diffusion results in decreasing surface roughness as presented for samples exposed in an inert atmosphere. The results of $R_a$ parameters differed for polished sample exposed in an inert gas at 700°C (Fig. 11) and recrystallized grains were visible on the macrophotographs of the samples (Fig. 10 a-b and d-e). However, no significant differences in fractal parameters describing surface roughness were observed for polished and ground samples (Table 5, Fig. 13a, b). It means that the temperature alone is not a determining factor causing the effect of decreasing roughness of initial copper boundary. However, for the sand-blasted sample (Fig. 13c), the higher the temperature of heat treatment was, the lowest value of surface roughness was obtained. It proves that the rate of surface diffusion of copper atoms is great enough for observation of changes in roughness for more convoluted surfaces, as the cold deformation, e.g. in grinding processes, increases the density of dislocations in the subsurface and introduces additional energy, which further increases the rate of outward diffusion of metallic atoms. Moreover, defects such as dislocations and grain boundaries make nucleation and growth of oxides on rough surfaces easier [37,38].

In the literature, there is a theory, that around each oxide island there is an active zone of oxygen capture, which radius is proportional to the oxygen surface diffusion rate [39,40] and that initial stages of oxidation is limited by oxygen surface diffusion [39]. After exposure at 700°C, the (110) clean, rough copper surface presented the higher density of occurrence of oxide islands, that also nucleate faster, even though grow laterally slower than oxide islands on smooth surfaces [5]. Enhanced surface roughness may hinder the mobility of oxygen surface diffusion, which causes higher density of oxide islands and consequently slows oxidation kinetics [41].

The above mentioned theory had been examined and verified in this study for samples oxidized at 700°C - polished sample presented the highest mass gain, while sand-blasted one – the lowest (Fig. 6). However, experimental results are in contradiction to the theory at 750°C – the highest mass gain was observed for sand-blasted sample and the lowest for polished one (Fig. 6). There are few explanations of such effect. The mobility of oxygen on the surface is greater at higher temperature, also connection of oxide islands is more favorable than creation of new nuclei [40,41]. It explains big, columnar grains of $Cu_2O$ on polished and ground samples (Fig. 7).

Behavior of sand-blasted sample at 750°C is substantially more complicated. First of all, the effect of temperature and diffusion of copper atoms as a result of high-temperature annealing is not enough for smoothening $Cu/Cu_2O$ interface to the level observed in Fig. 8f, which means that such effect is possible due mostly to chemical reactions between copper and oxygen. Moreover, as the theory says, the increased roughness may hinder surface oxidation and addition of new oxygen atoms to oxide islands. Instead, new oxide nuclei are formed, which result in creation of many small oxide grains – exactly as it is visible in Fig. 7f. Because sand-blasted sample is characterized by the greatest surface roughness and deformation, the density of grain boundaries and dislocations is also the greatest within it, which in addition to greater mobility of atoms at higher temperature results in greatest mass gain for this sample.

The issues that are connected with nucleation of new oxide islands and their growth are complicated and require extensive studies using sophisticated methods such as transmission electron microscopy (TEM) and its derivatives or scanning tunneling microscopy (STM) that provide atomistic information about behavior of copper at high temperature. Although fractal analysis is not a perfect method it is also simple and cheap and enables for visual and quantitative demonstration of changes that appear on the initial copper boundary and therefore for better understanding of surface reactions that may lead for further, more efficient research.

**5. Conclusions.**

1. Results of surface roughness of copper samples after polishing, grinding and sand-blasting processes measured by contact profilometer and evaluated by fractal analysis revealed similar trends: polished sample is characterized by the lowest values of roughness parameters and sand-blasted by the highest. The differences in surface roughness between polished and ground samples is much greater while measuring it by contact profilometer than basing on fractal analysis of binary scale images.

2. After oxidizing of polished and ground samples at 700°C, roughness of $Cu/Cu_2O$ boundary is higher than before heat treatment. However, the same prepared samples after exposure at 750°C presented the lowest value of roughness parameters. For sand-blasted samples increase in temperature resulted in the lowest roughness.

3. No significant differences in surface roughness of copper were observed for polished and ground samples after exposure in an inert gas at 700 and 750°C. For sand-blasted sample, important differences were noted – the higher the temperature of heat treatment was, the lowest value of surface roughness was obtained.

4. In case of sand-blasted sample two phenomena are responsible for smoothening of $Cu/Cu_2O$ boundary – chemical reaction between copper and oxygen and surface diffusion of copper due to high temperature annealing. The mobility of atoms at higher temperature is greater, so that the effect is substantially more visible at 750°C.

5. Surface diffusion of copper atoms has very little contribution in changing surface roughness of Cu/Cu$_2$O boundary for polished and ground samples. Smoothening of the initial boundary of cooper substrate for these samples is caused mainly by chemical reaction between copper and oxygen in oxidation process.

6. Fractal analysis seems to be a useful tool for evaluation of surface roughness of surfaces that are unavailable for measuring while using conventional contact profilometer. The method provides also many information on complexity of surface, which may lead for better understanding of materials' behavior at high temperature .

**Acknowledgement**

This research was financed within the Marie Curie COFUND scheme and POLONEZ program from the National Science Centre, Poland. POLONEZ Grant No. 2015/19/P/ST8/03995. This project has received funding from the European Union's Horizon 2020 research and innovation programme under the Marie Skłodowska-Curie Grant Agreement No. 665778.

**References**

[1] R. H. Havemann, J. A. Hutchby, High-performance interconnects: an integration overview, Proc. IEEE. 89 (2001) 586–601. doi:10.1109/5.929646.

[2] K. Fujita, D. Ando, M. Uchikoshi, K. Mimura, M. Isshiki, New model for low-temperature oxidation of copper single crystal, Appl. Surf. Sci. 276 (2013) 347–358. doi:https://doi.org/10.1016/j.apsusc.2013.03.096.

[3] E. Touzé, C. Cougnon, Study of the air-formed oxide layer at the copper surface and its impact on the copper corrosion in an aggressive chloride medium, Electrochim. Acta. 262 (2018) 206–213. doi:https://doi.org/10.1016/j.electacta.2017.12.187.

[4] K. Han, M. Tao, Electrochemically deposited p–n homojunction cuprous oxide solar

cells, Sol. Energy Mater. Sol. Cells. 93 (2009) 153–157. doi:https://doi.org/10.1016/j.solmat.2008.09.023.

[5] C. Gattinoni, A. Michaelides, Atomistic details of oxide surfaces and surface oxidation: the example of copper and its oxides, Surf. Sci. Rep. 70 (2015) 424–447. doi:https://doi.org/10.1016/j.surfrep.2015.07.001.

[6] B. Maack, N. Nilius, Oxidation of polycrystalline copper films – Pressure and temperature dependence, Thin Solid Films. 651 (2018) 24–30. doi:https://doi.org/10.1016/j.tsf.2018.02.007.

[7] P. Kofstad, High Temperature Corrosion, Elsevier Applied Science, London/New York, 1988.

[8] D.J. Young, High Temperature Oxidation and Corrosion of Metals, Elsevier Science, 2016. https://books.google.pl/books?id=TVXBBwAAQBAJ.

[9] E.A. Goldstein, T.M. Gür, R.E. Mitchell, Modeling defect transport during Cu oxidation, Corros. Sci. 99 (2015) 53–65. doi:https://doi.org/10.1016/j.corsci.2015.05.067.

[10] S. Mrowec, Podstawy teorii utleniania metali i stopów, Wydawnictwa Naukowo-Techniczne, Warszawa, 1964.

[11] Z. Grzesik, M. Migdalska, Oxidation Mechanism of Cu2O and Defect Structure of CuO at High Temperatures, High Temp. Mater. Process. 30 (2011). doi:10.1515/htmp.2011.046.

[12] S. Mrowec, A. Stokłosa, Oxidation of copper at high temperatures, Oxid. Met. 3 (1971) 291–311. doi:10.1007/BF00603530.

[13] Y. Zhu, K. Mimura, M. Isshiki, Oxidation Mechanism of Copper at 623-1073 K, Mater. Trans. - MATER TRANS. 43 (2002) 2173–2176. doi:10.2320/matertrans.43.2173.

[14] K. Mimura, L. Jae Won, M. Isshiki, Y. Zhu, Q. Jiang, Brief review of oxidation kinetics of copper at 350 °C to 1050 °C, 2006. doi:10.1007/s11661-006-1074-y.

[15] L. Jinlong, Y. Meng, H. Miura, L. Tongxiang, The effect of surface enriched chromium and grain refinement by ball milling on corrosion resistance of 316L stainless steel, Mater. Res. Bull. 91 (2017) 91–97. doi:https://doi.org/10.1016/j.materresbull.2017.03.022.

[16] H. Pei, Z. Wen, Z. Li, Y. Zhang, Z. Yue, Influence of surface roughness on the oxidation behavior of a Ni-4.0Cr-5.7Al single crystal superalloy, Appl. Surf. Sci. 440 (2018) 790–803. doi:https://doi.org/10.1016/j.apsusc.2018.01.226.

[17] L. Yuan, X. Chen, S. Maganty, J. Cho, C. Ke, G. Zhou, Enhancing the oxidation resistance of copper by using sandblasted copper surfaces, Appl. Surf. Sci. 357 (2015) 2160–2168. doi:https://doi.org/10.1016/j.apsusc.2015.09.203.

[18] D. Serafin, W.J. Nowak, B. Wierzba, The effect of surface preparation on high temperature oxidation of Ni, Cu and Ni-Cu alloy, Appl. Surf. Sci. 476 (2019) 442–451. doi:https://doi.org/10.1016/j.apsusc.2019.01.122.

[19] R. Panat, K. Jimmy Hsia, D. Cahill, Evolution of surface waviness in thin films via volume and surface diffusion, J. Appl. Phys. 97 (2004). doi:10.1063/1.1827920.

[20] N. Sheng, K. Horke, A. Meyer, M.R. Gotterbarm, R. Rettig, R.F. Singer, Surface recrystallization and its effect on oxidation of superalloy C263, Corros. Sci. 128 (2017) 186–197. doi:https://doi.org/10.1016/j.corsci.2017.09.020.

[21] R. Kromer, S. Costil, C. Verdy, S. Gojon, H. Liao, Laser surface texturing to enhance adhesion bond strength of spray coatings – Cold spraying, wire-arc spraying, and atmospheric plasma spraying, Surf. Coatings Technol. 352 (2018) 642–653. doi:https://doi.org/10.1016/j.surfcoat.2017.05.007.

[22] R. Eriksson, S. Sjöström, H. Brodin, S. Johansson, L. Östergren, X.-H. Li, TBC bond coat–top coat interface roughness: Influence on fatigue life and modelling aspects, Surf. Coatings Technol. 236 (2013) 230–238. doi:https://doi.org/10.1016/j.surfcoat.2013.09.051.

[23] M. Gupta, R. Eriksson, U. Sand, P. Nylén, A diffusion-based oxide layer growth model using real interface roughness in thermal barrier coatings for lifetime assessment, Surf. Coatings Technol. 271 (2015) 181–191. doi:https://doi.org/10.1016/j.surfcoat.2014.12.043.

[24] A.M. Huntz, B. Lefevre, F. Cassino, Roughness and oxidation: application to NiO growth on Ni at 800°C, Mater. Sci. Eng. A. 290 (2000) 190–197. doi:https://doi.org/10.1016/S0921-5093(00)00944-8.

[25] S. Chen, F. Ke, M. Zhou, Y. Bai, Atomistic investigation of the effects of temperature and surface roughness on diffusion bonding between Cu and Al, Acta Mater. 55 (2007) 3169–3175. doi:https://doi.org/10.1016/j.actamat.2006.12.040.

[26] Z. Liu, C. Patzig, S. Selle, T. Höche, P. Gumbsch, C. Greiner, Stages in the tribologically-induced oxidation of high-purity copper, Scr. Mater. 153 (2018) 114–117. doi:https://doi.org/10.1016/j.scriptamat.2018.05.008.

[27] S.M. Poljacek, D. Risovic, K. Furic, M. Gojo, Comparison of fractal and profilometric methods for surface topography characterization, Appl. Surf. Sci. 254 (2008) 3449–

3458. doi:https://doi.org/10.1016/j.apsusc.2007.11.040.

[28] A. Roos, M. Bergkvist, C.-G. Ribbing, J.M. Bennett, Quantitative interface roughness studies of copper oxide on copper, Thin Solid Films. 164 (1988) 5–11. doi:https://doi.org/10.1016/0040-6090(88)90101-0.

[29] V. Hotař, Fractal geometry for industrial data evaluation, Comput. Math. with Appl. 66 (2013) 113–121. doi:https://doi.org/10.1016/j.camwa.2013.01.015.

[30] V. Hotař, A. Hotař, Fractal dimension used for evaluation of oxidation behaviour of Fe-Al-Cr-Zr-C alloys, Corros. Sci. 133 (2018) 141–149. doi:https://doi.org/10.1016/j.corsci.2018.01.017.

[31] E.S. Gadelmawla, M.M. Koura, T.M.A. Maksoud, I.M. Elewa, H.H. Soliman, Roughness parameters, J. Mater. Process. Technol. 123 (2002) 133–145. doi:https://doi.org/10.1016/S0924-0136(02)00060-2.

[32] M. Papanikolaou, K. Salonitis, Fractal roughness effects on nanoscale grinding, Appl. Surf. Sci. 467–468 (2019) 309–319. doi:https://doi.org/10.1016/j.apsusc.2018.10.144.

[33] C.A. Brown, W.A. Johnsen, R.M. Butland, J. Bryan, Scale-Sensitive Fractal Analysis of Turned Surfaces, CIRP Ann. 45 (1996) 515–518. doi:https://doi.org/10.1016/S0007-8506(07)63114-X.

[34] J.-J. Park, S.-I. Pyun, Pit formation and growth of alloy 600 in Cl− ion-containing thiosulphate solution at temperatures 298–573 K using fractal geometry, Corros. Sci. 45 (2003) 995–1010. doi:https://doi.org/10.1016/S0010-938X(02)00212-3.

[35] W. Nowak, D. Naumenko, G. Mor, F. Mor, D.E. Mack, R. Vassen, L. Singheiser, W.J. Quadakkers, Effect of processing parameters on MCrAlY bondcoat roughness and lifetime of APS–TBC systems, Surf. Coatings Technol. 260 (2014) 82–89. doi:https://doi.org/10.1016/j.surfcoat.2014.06.075.

[36] Sfrax 1.0., (n.d.). http://www.surfract.com/sfrax.html (accessed September 24, 2018).

[37] D. Pradhan, G.S. Mahobia, K. Chattopadhyay, V. Singh, Effect of surface roughness on corrosion behavior of the superalloy IN718 in simulated marine environment, J. Alloys Compd. 740 (2018) 250–263. doi:https://doi.org/10.1016/j.jallcom.2018.01.042.

[38] M.J. Seo, H.-S. Shim, K.M. Kim, S.-I. Hong, D.H. Hur, Influence of surface roughness on the corrosion behavior of Alloy 690TT in PWR primary water, Nucl. Eng. Des. 280 (2014) 62–68. doi:https://doi.org/10.1016/j.nucengdes.2014.08.023.

[39] J. C Yang, M. Yeadon, B. Kolasa, J. M Gibson, The Homogeneous Nucleation Mechanism of Cu 2O on Cu(001), Scr. Mater. - Scr. MATER. 38 (1998) 1237–1242. doi:10.1016/S1359-6462(98)00026-8.

[40] J.C. Yang, G. Zhou, In situ ultra-high vacuum transmission electron microscopy studies of the transient oxidation stage of Cu and Cu alloy thin films, Micron. 43 (2012) 1195–1210. doi:https://doi.org/10.1016/j.micron.2012.02.007.

[41] G. Zhou, J.C. Yang, In situ UHV-TEM investigation of the kinetics of initial stages of oxidation on the roughened Cu(110) surface, Surf. Sci. 559 (2004) 100–110. doi:https://doi.org/10.1016/j.susc.2004.04.046.